\documentclass[10pt, aps, prl, amsmath, twocolumn, showpacs, superscriptaddress]{revtex4-1}
\usepackage{amsmath}
\usepackage{amssymb}
\usepackage{mathtools}
\usepackage{esint}

\usepackage{color}
\definecolor{dgreen}{rgb}{0,0.7,0}

\newcommand{\ee}{\mathrm{e}}

\newcommand{\al}{\alpha}
\newcommand{\ep}{\epsilon}

\begin{document}
\title{Exact extremal statistics in the classical $1d$ Coulomb gas}
\author{Abhishek Dhar}
\affiliation{International Centre for Theoretical Sciences, TIFR, Bangalore 560089, India}
\author{Anupam Kundu}
\affiliation{International Centre for Theoretical Sciences, TIFR, Bangalore 560089, India}
\author{Satya N. Majumdar}
\affiliation{LPTMS, CNRS, Univ. Paris-Sud, Universit\'{e} Paris-Saclay, 91405 Orsay, France}
\author{Sanjib Sabhapandit}
\affiliation{Raman Research Institute, Bangalore 560080, India}
\author{Gr\'egory Schehr}
\affiliation{LPTMS, CNRS, Univ. Paris-Sud, Universit\'{e} Paris-Saclay, 91405 Orsay, France}
\date{\today}

\begin{abstract}
We consider a one-dimensional classical Coulomb gas of $N$ like-charges in a harmonic potential --~also known as the one-dimensional one-component plasma (1dOCP). We compute analytically the probability distribution of the position $x_{\max}$ of the rightmost charge in the limit of large $N$. We show that the typical fluctuations of $x_{\max}$ around its mean are described by a non-trivial scaling function, with asymmetric tails. This distribution is different from the Tracy-Widom distribution of $x_{\max}$ for the Dyson's log-gas. We also compute the large deviation functions of $x_{\max}$ explicitly and show that the system exhibits a third-order phase transition, as in the log-gas. Our theoretical predictions are verified numerically.    
\end{abstract}


\pacs{}
\maketitle


The Tracy-Widom (TW) distribution has emerged ubiquitously in diverse systems in the recent past \cite{buchanan,quanta}. It was originally discovered as the limiting distribution of 
the top eigenvalue $x_{\max}$ of an $N \times N$ Gaussian random matrix~\cite{Tracy94}. Since then, it has appeared in various areas of  physics \cite{Satya_lectures,Krug2010}, mathematics \cite{BDJ1999,Joh2001}, and information theory \cite{Moustakas}. For example, in physics it has appeared in stochastic growth models and related directed polymer in $1+1$ dimensional random media belonging to the Kardar-Parisi-Zhang (KPZ) universality class \cite{Joha2000,PS2000,MS2004,SS2010,CDR2010,Dot2010,ACQ2011}, non-intersecting Brownian motions \cite{FMS2011}, non-interacting fermions in a one-dimensional trapping potential \cite{Eisler2013,Dean15,Dean16}, disordered mesoscopic systems \cite{Beenaker} and even in the Yang-Mills gauge theory in $2$-dimensions \cite{FMS2011}. It has also been measured experimentally in several systems including liquid crystals \cite{Takeuchi}, coupled fiber lasers \cite{Davidson} or disordered superconductors \cite{Lem2013}. The TW distribution describes the probability of {\it typical} fluctuations of $x_{\max}$ around its mean. In contrast, the {\it atypical} fluctuations of $x_{\max}$ to the left and right, far from its mean, are described respectively by the left and right large deviation tails. These tails have been computed explicitly \cite{BDG2001,Dean06,Dean08,Majumdar09} and shown to correspond to two different thermodynamic phases separated by a third order phase transition \cite{Nadal,Majumdar14}. Similar third order phase transitions have also been found in a variety of other systems \cite{SMCF,CP2013,Majumdar14,ATW2014,Ledou16,CMV2016,Baruch17}.

%
\begin{figure}[t]
  \centering
      \includegraphics[scale=0.3]{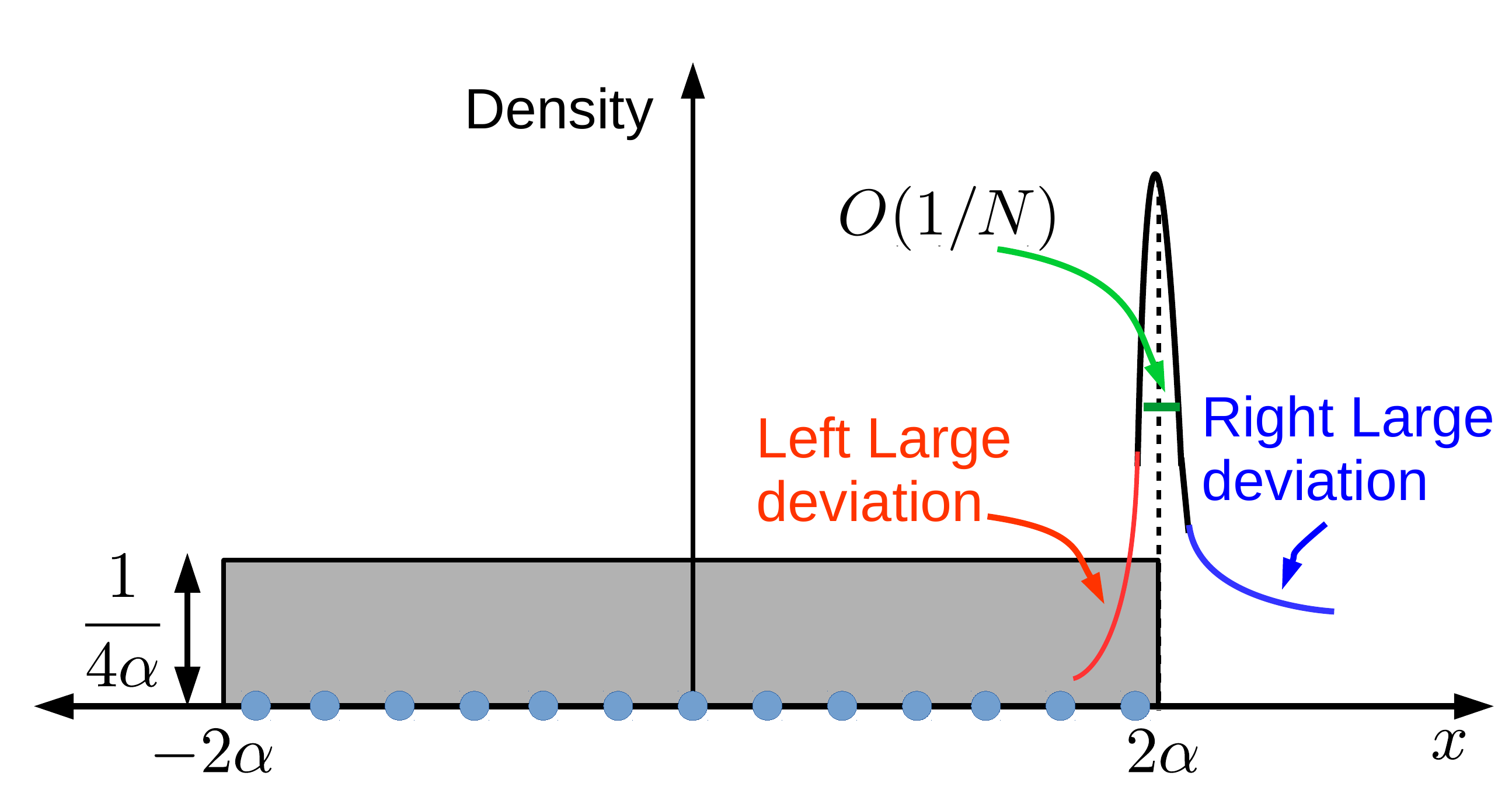}
  \caption{Schematic plot of the PDF of $x_{\max}$ with a peak around the right edge $2\al$ of the average density profile. The typical fluctuations (black) of ${O}(1/N)$ are described by $F_\alpha'(x)$ [see \eqref{limit.1}], while the large deviations of ${O}(1)$ to the left and right of the mean are described by the left (red) and right (blue) large deviation tails.}
\label{fig1}
\end{figure}


For Gaussian ensembles in random matrix theory (RMT), the joint probability distribution function (PDF) of the $N$ real eigenvalues $\{x_1, \cdots, x_N \}$ is known~explicitly~\cite{Mehta91,Forrester10}
\begin{equation}
\mathcal{P}(\{x_i\})= B_N~\ee^ {-\frac{\beta}{2}\left( N \sum_{i=1}^N x_i^2 - \sum_{i \neq j}\log (|x_i-x_j|)\right) }, \label{JPDF-rm}
\end{equation}
where $B_N$ is a normalisation constant and $\beta=1$, $2$ and $4$ depending on the symmetries of the matrices~\cite{Mehta91, Forrester10}. This joint PDF can be interpreted as the equilibrium Gibbs distribution of a gas of $N$ charges with positions $x_i$'s that are confined on a line in the presence of an external harmonic potential and repelling each other via two dimensional logarithmic Coulomb interactions. This system is known as Dyson's log-gas~\cite{Dyson62}. In this picture the largest eigenvalue $x_{\max}=\max \{x_1, \cdots,x_N\}$ corresponds to the position of the rightmost particle. The average density of eigenvalues $\rho_N(x)$ converges for large $N$ to the Wigner semi-circular law, $\rho_\infty(x) \sim \sqrt{2-x^2}/\pi$ which has a finite support $[-\sqrt{2},+\sqrt{2}]$. Hence the average $\langle x_{\max}\rangle \sim \sqrt{2}$ for large $N$. The typical fluctuations of $x_{\max}$ around its mean $\sqrt{2}$ are of ${O}(N^{-2/3})$. On this scale the cumulative distribution $Q(w,N) = {\rm Prob}(x_{\max}\leq w,N)$, takes the scaling form 
\begin{eqnarray}
Q(w,N) \approx {\cal F}_\beta\left(\sqrt{2}N^{2/3}(w-\sqrt{2})\right) \label{TW} \;,
\end{eqnarray}
where ${\cal F}_\beta(x)$ is the Tracy-Widom distribution. This scaling function can be written in terms of the solution of a Painlev\'e II equation~\cite{Tracy94} and has non-Gaussian tails. Interestingly, even though the TW distribution was derived originally for a harmonic confining potential, it has been shown to be universal with respect to the shape of the confining potential, as long as the average density vanishes at the upper edge as a square root (as in the case of the Wigner semi-circular law).  A natural question then arises whether the TW distribution for $x_{\max}$ is robust when one changes, instead of the confining potential, the nature of the repulsive interaction between the charges.



A natural candidate model to address this question is the system of one-dimensional charges in a harmonic potential but interacting pairwise via the true $1d$ Coulomb potential. The energy of this system is given by
\begin{equation}
E(\{x_i\})=\frac{N^2}{2} \sum_{i=1}^N x_i^2 - \alpha N \sum_{i \neq j} |x_i-x_j| \;, \label{E-1Dc}
\end{equation}
where $\alpha \geq 0$ denotes the strength of the Coulomb repulsion. The choice of the powers of $N$ in the coupling constants is such that for large $N$ (i) both terms in the energy are of same order and (ii) the charges are confined in a region whose span is $O(1)$. Indeed this is also the 1dOCP or ``jellium'' model, where $N$ charges of the same sign interact in the presence of a uniform background of opposite charges, assuring charge neutrality.   This model is a paradigm for $1d$ charged plasma \cite{Choquard1981} as several observables can be  calculated analytically~\cite{Lenard61,Prager62,Baxter63,Dean2010,TT2015}. 

In this Letter, we show that the statistics of the position of the rightmost charge $x_{\max}$ can also be computed exactly in this 1dOCP model. Our main result is to show that the limiting distribution of the typical fluctuations of $x_{\max}$ in this model is indeed different from the TW distribution, found in the Dyson's log-gas. Moreover, by computing the left and the right large deviation functions explicitly, we find that the third-order phase transition between a pushed gas (left large deviation) and a pulled gas (right large deviation) is still present in the 1dOCP model, as in the case of the log-gas. Interestingly, a similar third-order phase transition between the pushed and the pulled phase was recently found \cite{Cunden2017} by analysing large deviation functions associated with the position of 
the farthest charge in a $d$-dimensional jellium model, though the limiting distribution of the position of the farthest charge is still open for this $d$-dimensional problem. In $d=1$ this corresponds to the distribution of the maximum of $|x_i|$'s of the charges.    

\begin{figure}[t]
  \centering
    \includegraphics[width=\linewidth]{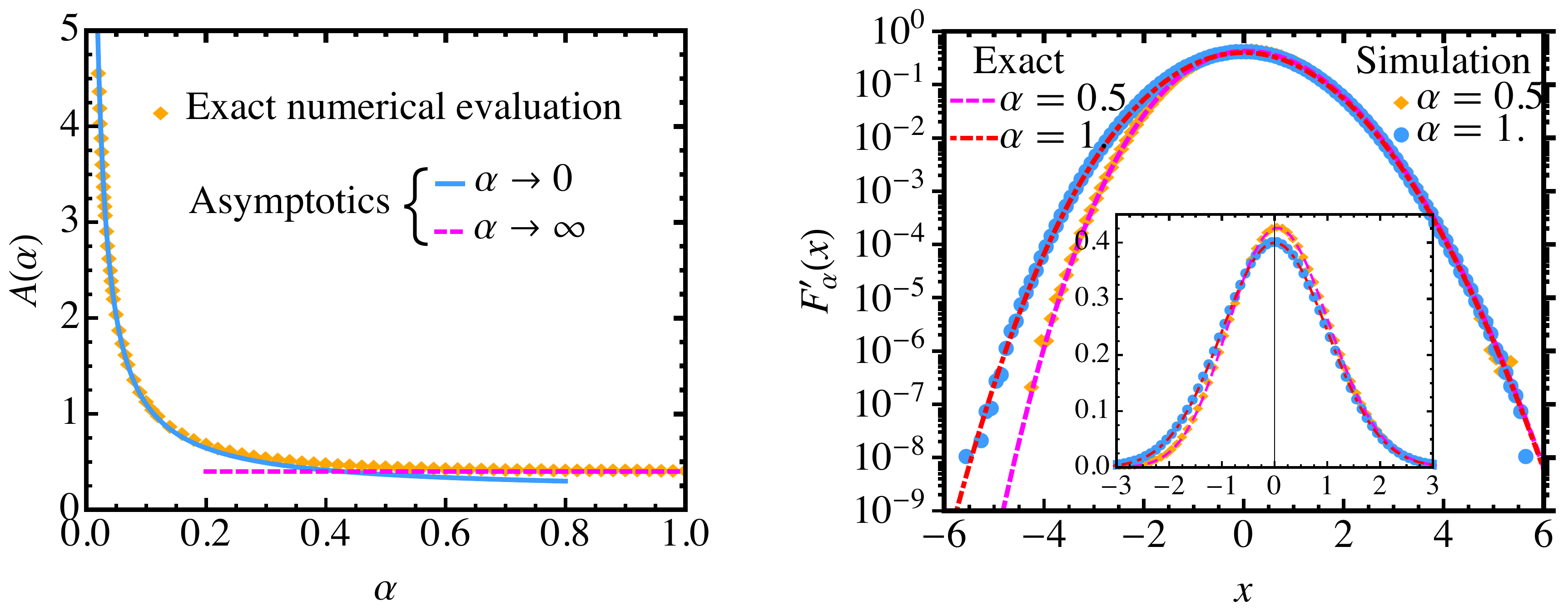}
  \caption{(Left): Plot of $A(\al)$ and numerical verification of its $\al \to 0$ and $\al \to \infty$ asymptotic. (Right): Comparison of the theoretical $F_\al'(x)$ obtained by solving numerically \eqref{limit.1} by a shooting method and $F_\al'(x)$ obtained from direct Monte-Carlo simulation of the 1dOCP (with $N=50$) for two different values of the coupling parameter $\alpha = 1$ and $\alpha = 0.5$. Inset shows the distribution in the normal scale.} 
\label{A-al-fig}
\end{figure}

We start with the joint PDF of the positions 
$x_i \in(-\infty, \infty)$ in the 1dOCP, given by the Boltzmann weight 
\begin{equation}
{\cal P}(\{x_i\})= \frac{1}{Z_N}~\exp[-E(\{x_i\})], \label{JPDF-1Dc}
\end{equation}
where $Z_N$ is the partition function and the energy $E(\{x_i\})$ is given in (\ref{E-1Dc}). In the large $N$ limit, the average density can be obtained by minimizing the energy $E(\{x_i\})$. It is easy to show that the minimum  
energy configuration is given by $x_j^*=\frac{2 \al}{N}(2j-N-1)$ ($j=1,\cdots, N$). Hence the particles are equi-spaced and the rightmost (leftmost) particle is at $x_N^*=2\al(1-1/N)$ (respectively at $x_{1}^*=-2\al(1-1/N)$). This implies that in the $N\to \infty$ limit, the average density profile $\rho_\infty(x)$ is flat: $\rho_{\infty}(x) = \frac{1}{4\al}$ for $-2\al \leq x \leq 2\al$~(see Fig. \ref{fig1}), in contrast to the Wigner semi-circle in the log-gas. Our focus here is on the large $N$ behavior of the cumulative distribution of the rightmost particle, 
\begin{equation}
Q(w,N) = \text{Prob}.[x_{\max} \leq w,N] \;. \label{Q_N} 
\end{equation}
To anticipate the scaling behavior of $Q(w,N)$, we first make the following observations. It follows from the above analysis of the average density that the mean position of the rightmost particle is at $\langle x_{\max}\rangle = x_N^* = 2 \alpha(1-1/N)$. Given that the average density is uniform, the typical separation between two adjacent particles is of $O(1/N)$ everywhere. Hence, the scale of typical fluctuations of $x_{\max}$ around its average is also of $O(1/N)$. This suggests that the cumulative probability distribution $Q(w,N)$, in the region of typical fluctuations where $|w-2\alpha| \sim O(1/N)$, should have the scaling form for large $N$, $Q(w,N) \approx F_\alpha(N(w-x_N^*))=F_\alpha(N(w-2\alpha) + 2 \alpha)$ where $F_\alpha(x)$ is a nontrivial scaling function (the analogue of the TW distribution for the log-gas). In this paper, we compute this scaling function $F_\alpha(x)$. In addition, for atypical fluctuations where $x_{\max} - \langle x_{\max} \rangle \sim O(1)$, both to the left and to the right of the mean, $Q(w,N)$ has large deviation tails that are also computed explicitly. More precisely, we find   
\begin{equation}
Q(w,N) \approx
\begin{cases}
\ee^{-N^3\, \Phi_{-}(w) + O(N^2)}  \, &\,0< 2\al-w \sim O(1) \\
F_\alpha\left[N(w-2\al)+2\al\right] \, &\, |2\al-w| \sim 
O(1/N)\\
1- e^{- N^2\, \Phi_+(w)+O(N)} \, &\, 0<w-2\al \sim O(1)
\end{cases}
\label{result-1}
\end{equation}
where $\Phi_-(w)$ and $\Phi_+(w)$ are the left and right rate functions. We show that the scaling function $F_\alpha(x)$ in the central regime satisfies 
a nonlocal eigenvalue equation 
\begin{equation}
\frac{dF_\alpha(x)}{dx}= A(\al)\, \ee^{-x^2/2}\, F_\alpha(x+4\al) \;,
\label{limit.1}
\end{equation}
with the boundary conditions: $F_\alpha(-\infty)=0$ and $F_\al(\infty)=1$. These boundary conditions, along with the fact that $F_\al(x) \geq 0$ for all $x$, uniquely fixes the eigenvalue $A(\al)$. Clearly, the scaling function $F_\al(x)$ is different from the TW distribution. 
While it is hard to compute $A(\alpha)$ explicitly for all $\alpha \geq 0$ (for a numerical plot of $A(\alpha)$, see Fig.~\ref{A-al-fig}), we can determine its small and large $\al$ behaviors: $A(\al) \to 1/(4\,e\,\al)$ as $\al \to 0$ and $A(\al) \to 1/\sqrt{2 \pi}$ as $\al \to \infty$.
%
From Eq. (\ref{limit.1}), we can derive the leading asymptotic tails of the PDF $F_\alpha'(x)$ for all $\alpha$
\begin{eqnarray}
F_\al'(x) \approx
\begin{cases}
&\exp\left[-|x|^3/{24\al}+ O(x^2)\right ] \; \textrm{as}\,\, x \to -\infty \\
& \vspace*{-0.25cm} \\
& \exp\left[-x^2/2 + O(x)\right] \hspace*{0.8cm}
\,\textrm{as}\,\, x \to \infty \;.
\end{cases}
\label{fx_asymp}
\end{eqnarray}   
We note that the leading $x \to -\infty$ behavior of $F_\al'(x)$ is identical to that of the TW distribution ${\cal F}'_{\beta=1/\alpha}(x)$, while the right tail of $F_\al'(x)$ decays faster than the TW right tail \cite{Dumaz2013}. This is indeed our main result. In addition, we also compute exactly the large deviations rate functions. For the left tail we find
\begin{eqnarray}\label{left_rf}
\Phi_{-}(w) = 
\begin{cases}
& \frac{(2\al-w)^3}{24\al}\;,\;\quad -2\al\leq w \leq 2\al \\
&Ê\vspace*{-0.3cm} \\
& \frac{w^2}{2} + \frac{2}{3} \al^2 \;, \;\quad\; w \leq -2\al \;.
\end{cases}
\end{eqnarray}
 For the right tail, we find
\begin{eqnarray}
\Phi_+(w)& =& \frac{(w-2\al)^2}{2}\;,\;\quad w>2\al \;. \label{right_rf} 
\end{eqnarray}
It is easy to check, using the asymptotic behavior of $F_\al'(x)$ for large $|x|$ in Eq. (\ref{fx_asymp}), that the central part of the distribution of $x_{\max}$ matches smoothly with the two large deviation regimes flanking this central part. Indeed, as discussed later, the vanishing of $\Phi_-(w)$ when $w \to 2\alpha$ as a cubic power 
is responsible for a third order phase transition at the critical point $w = 2 \alpha$, in very much the same way as in the log-gas \cite{Majumdar14}.

We start from the joint PDF of $\{x_i\}$'s in Eq. \eqref{JPDF-1Dc}. We note that $Q(w,N) = \text{Prob}.[x_{\max} \leq w] = \text{Prob.}(x_1\leq w, \cdots,x_N\leq w)$. Hence it can be expressed as the ratio of two partition functions  
\begin{eqnarray}
Q(w,N)&=& \frac{Z_N(w)}{Z_N(\infty)},~~\text{where,}\label{Q_N-1} \\
Z_N(w)&=& \int_{-\infty}^w dx_1\cdots \int_{-\infty}^w dx_N~\ee^{-E(\{x_i\})}, \label{Z_N-w}
\end{eqnarray}
with $E(\{x_i\})$ given in \eqref{E-1Dc} and we have suppressed the $\alpha$-dependence in $Z_N(w)$ for simplicity. Note that $Z_N(w)$ can be interpreted as the partition function of the 1dOCP in presence of a hard wall at $w$. Below, we analyse $Q(w,N)$ in the central regime first, followed by the two large deviation tails.  

{\it Central regime:} Noting that the energy function $E(\{x_i\})$ in \eqref{E-1Dc} is symmetric under permutations over the positions $(x_1,x_2,\cdots,x_N)$, we write
\begin{eqnarray}
Z_N(w) &=&N! \prod_{k=1}^N\int_{-\infty}^w dx_k~\ee^{-E(\{x_k\})}\prod_{j=2}^N\Theta(x_j-x_{j-1}), ~~~~~\label{Z_N-w-ord}
\end{eqnarray}
where $\Theta(x)$ is the Heaviside theta function. For an ordered configuration $(x_1 < x_2< \cdots <x_N)$, one can eliminate the absolute values $|x_i - x_j|$ 's and rewrite the energy function $E(\{x_i\})$ in \eqref{E-1Dc}  as 
$E(\{x_i\})=\frac{N^2}{2} \sum_{i=1}^N \left( x_i - \frac{2 \alpha}{ N}( 2i-N-1)\right)^2 + C_N(\alpha)$, where $C_N(\alpha)= 2 \alpha^2 \sum_{i=1}^N \left( 2i-N-1\right)^2$ is just a constant. This trick of eliminating the absolute values via ordering has been used before for 1dOCP in numerous contexts \cite{Lenard61,Baxter63,Dean2010,TT2015}. Performing a change of variables $\ep_k=\left(Nx_k-2\al(2k-N-1)\right)$ for all $k=1,2,\cdots,N$ in \eqref{Z_N-w-ord}, we can rewrite $Z_N(w) = N!~ D_\alpha\left(N\left(w-\frac{2 \al}{N}(N-1)\right),N\right)$ where 
\begin{eqnarray}
D_\al(x,N) =&&\int_{-\infty}^x d \ep_N\int_{-\infty}^{\ep_N+4 \al}d \ep_{N-1}\dots \int_{-\infty}^{\ep_2+4\al} d\ep_1~\nonumber\\
&&\times \,\ee^{-\frac{1}{2}\sum_{i=1}^N \ep_i^2}  \label{D_N} \;.
\end{eqnarray}
Therefore setting $x = N\left(w-\frac{2 \al}{N}(N-1)\right)$, $Q(w,N)$ in Eq.~(\ref{Q_N-1}) can be written as
\begin{eqnarray}
Q(w,N) = \frac{D_\al(x,N)}{D_\al(\infty,N)} \equiv F_\alpha(x,N) \;. \label{def_FN}
\end{eqnarray}
Taking derivative with respect to $x$ in Eq. (\ref{def_FN}), and using Eq. (\ref{D_N}), we obtain
\begin{equation}\label{dF_N.1}
\frac{d\,F_\al(x,N)}{d\,x} = \frac{D_{\al}(\infty,N-1)}{D_{\al}(\infty,N)}\,\ee^{-\frac{x^2}{2}} F_{\al}(x+4\al,N-1) \;.
\end{equation}
To estimate the ratio ${D_{\al}(\infty,N-1)}/{D_{\al}(\infty,N)}$ for large $N$, we note from Eq. (\ref{D_N}) that $D_\al(\infty,N)$ can be interpreted as the partition function of an auxiliary gas of particles with positions $\xi_1, \xi_2, \cdots, \xi_N$ confined in an external harmonic potential and with the one-sided constraint $\xi_{k-1}<\xi_k + 4 \al$ for all $k=2,3,\cdots N$. Indeed this constraint provides a short-range interaction between the particles. Thus our original problem of the 1dOCP which has long-range interaction is mapped onto a problem of short-ranged interacting particles. For such a short-ranged system, it is natural to expect that the free energy is extensive in $N$. Thus one would expect that, for large $N$, the partition function must scale as $D_\al(\infty,N) \sim [A(\alpha)]^{-N}$ where $\ln A(\alpha)$ is the free energy per particle. Thus the ratio ${D_{\al}(\infty,N-1)}/{D_{\al}(\infty,N)} \to A(\alpha)$ as $N \to \infty$. This suggests that $F_\al(x,N)$ should converge to a limiting form $F_\al(x)$ for large $N$, which then satisfies the nonlocal eigenvalue equation (\ref{limit.1}). Thus the eigenvalue $A(\alpha)$ has a physical interpretation:  $\ln A(\alpha)$ is the free energy per particle of a short-ranged interacting gas. However, computing analytically $A(\alpha)$ for all $\alpha$ seems hard. Interestingly, Baxter \cite{Baxter63} encountered a similar nonlocal eigenvalue equation while computing the partition function of the 1dOCP in a finite box $[-L,+L]$ and analysed the eigenvalue $A(\alpha)$ in the two limits $\alpha \to 0$ and $\alpha \to \infty$. Translating his results to our problem, following a simple rescaling of the parameters, we obtain the asymptotic results for $A(\alpha)$ announced before. 

It is straightforward to derive the asymptotic tails of $F_\alpha'(x)$ in Eq. (\ref{fx_asymp}). We consider first the limit $x \to \infty$ where $F_\alpha(x + 4 \alpha) \to 1$ on the right hand side (rhs) of Eq.~(\ref{fx_asymp}). Hence, to leading order, $F_\alpha'(x) \approx A(\al)\,\ee^{-x^2/2}$, providing the Gaussian right tail in Eq. (\ref{fx_asymp}). To compute the left tail, we make a simple ansatz that $F_\alpha(x) \approx \ee^{-a_0\,|x|^\delta}$ as $x \to -\infty$, where $a_0$ and $\delta$ are to be determined. Substituting this ansatz in the rhs of Eq. (\ref{fx_asymp}) yields $\approx A(\al)\,\ee^{-x^2/2-a_0\,(|x|-4\al)^\delta} $. For large $|x|$, $(|x|-4\al)^\delta \sim |x|^\delta(1 - 4\,\alpha\,\delta/|x|)$ to leading orders. Hence the rhs behaves as $A(\al) \ee^{-a_0\,|x|^\delta - x^2/2 + 4\,\al\,\delta\,|x|^{\delta-1}}$. The left hand side (lhs) of Eq. (\ref{fx_asymp}) behaves as $\approx \ee^{-a_0\, |x|^\delta}$ to leading order. Comparing both sides, we see that the term $x^2/2$ and $|x|^{\delta-1}$ on the rhs must cancel each other implying $\delta = 3$ and $a_0 = 1/(8\alpha \,\delta) = 1/(24\, \alpha)$. This provides the leading left tail of $F_\alpha'(x)$ in Eq. (\ref{fx_asymp}).

%
For general $\al>0$, determining explicitly the eigenvalue $A(\alpha)$ and the full scaling function $F_\alpha(x)$ is difficult. However they can be obtained by solving \eqref{limit.1} numerically by tuning the value of $A(\al)$ using the standard shooting method. This gives $F_\al(x)$ and $A(\al)$ simultaneously. In Fig.~\ref{A-al-fig} (left panel), we plot $A(\al)$ {vs.} $\al$ and compare with its predicted asymptotics. In Fig.~\ref{A-al-fig} (right panel), we compare $F_\al'(x)$ evaluated numerically from this shooting method, with the one obtained from direct Monte-Carlo simulation of the 1dOCP. The agreement is excellent.

 {\it Left large deviation function.} We consider $Q(w,N)$ in Eqs. (\ref{Q_N-1}) and (\ref{Z_N-w}) with $0<2 \al-w \sim O(1)$. Since $w$ represents the position of the hard wall, $w<2\al$ corresponds to ``pushing'' the charges to the left of the right edge at $2\al$. This disturbs the originally flat density and leads to a collective reorganisation of all the charges, as in the case of the log-gas \cite{Dean06,Dean08}. We get instead a new equilibrium density that minimizes the energy, i.e., a new saddle point of the integral in Eq. (\ref{Z_N-w}). It is well known that in the jellium model, the bulk density is insensitive to the location of a wall~\cite{ASZ2014}. This implies that in the bulk, the density is still given by the original equilibrium value $1/(4\al)$, for $-2\al < w \leq 2 \al$. Hence, when the wall moves to the left of $2\al$, all the charges that get pushed by the wall must get absorbed {\it at} the wall. This observation leads us to look for a saddle point density of the form 
\begin{eqnarray}\label{rho_w}
\rho_w(x) = \frac{1}{4 \al} + C\, \delta(x-w) \;, \quad -B \leq x \leq w \;,
\end{eqnarray}   
where the constant bulk density is supported over the interval $[-B,w]$. We then minimize the energy with respect to the two variational parameters $B$ and $C$. Skipping details (see \cite{supp_mat}), we find that 
\begin{eqnarray}\label{AB}
C = {1}/{2}-{w}/{(4\al)} \;, \;\quad B = 2 \al \;,
\end{eqnarray}  
as long as $-2\al\leq w\leq 2 \al$. When $w$ hits $-2\al$ from the right, all the charges get absorbed at the wall $w$ and the saddle point density is just $\rho_w(x) = \delta(x-w)$, for all $w \leq -2\al$. Substituting $\rho_w(x)$ in the energy \cite{supp_mat}, we find the results for $\Phi_-(w)$ given in Eq. (\ref{left_rf}).


 {\it Right large deviations.} For fluctuations $(x_{\max}-2\al) \sim {O}(1)$ to the right of the edge $2 \al$, we consider the PDF $\partial_w Q(w,N)$ in Eqs. (\ref{Q_N-1}) and (\ref{Z_N-w}) with $w>2\al$. It turns out that the configuration that dominates this integral is one where the rightmost charge is at $w > 2 \al$, while the rest of the $N-1$ charges stay in the equilibrium configuration with a flat profile over the interval $[-2\al,+2\al]$. This is analogous to the ``pulled'' phase in the log-gas \cite{Majumdar09}. Thus, for large $N$, the PDF can be approximated as $\partial_w Q(w,N) \approx \ee^{-\Delta E_{\rm pulled}}$, where $\Delta E_{\rm pulled}$ is the energy cost of pulling the rightmost particle from the ``sea'' of $N-1$ particles in the equilibrium flat configuration. This energy cost can be estimated from Eq.~(\ref{E-1Dc}): a first contribution from the change in the external potential energy of the rightmost charge and a second due to the interaction of the rightmost charge with the $(N-1)$ other particles. One gets (see \cite{supp_mat} for details): $\Delta\,E_{\rm pulled} \approx N^2\left( \frac{w^2}{2} - \frac{1}{2}\int_{-2\al}^{2\al} (w-x) \,dx\right)$ up to a constant. This gives $\partial_wQ_N(w) \approx \ee^{-N^2 \Phi_+(w)}$ where $\Phi_+(w)$ is given in~(\ref{right_rf}).    

Since $Q(w,N)$ is the ratio of two partition functions (\ref{Z_N-w}), $-\ln Q(w,N)$ is the free energy difference between the pushed (left) and the pulled (right) phase. From Eq.~(\ref{result-1}), this free energy $\propto \Phi_-(w)$ has a singular behavior at the critical point $w=2\al$. Indeed it vanishes as a cubic power as $w\to 2\al$ from the left [see Eq. (\ref{left_rf})], leading to a discontinuity of the third derivative of $\Phi_-(w)$ at $w = 2\al$. This third order phase transition at $w = 2\al$ is similar to the one in the log-gas. Unlike in the log-gas, there is an additional third-order phase transition in this 1dOCP when $w \to -2\al$ [see Eq.~(\ref{left_rf})]. However, this transition is not of the ``pushed-pulled'' type like the one at $w = 2\al$, but rather a condensation-type transition as all charges accumulate at the wall for $w\leq -2\al$.

{\it Conclusion:} In this Letter we have studied analytically the distribution of the position of the rightmost particle $x_{\max}$ of a $1d$ Coulomb gas confined in an external harmonic potential (1dOCP), in the limit of large number of particles $N$. We have obtained the limiting large $N$ distribution describing the typical fluctuations of $x_{\max}$ around its mean, and shown that it is different from the Tracy-Widom distribution of the log-gas. We also computed the rate functions associated with atypically large fluctuations around the mean and found a third order phase transition between a pushed and a pulled phase, as in the log-gas. Our work raises several interesting questions. For instance, how universal is the limiting distribution of $x_{\max}$if one changes the confining potential or the pairwise repulsive interaction? It would be challenging to study $x_{\max}$ with a repulsive interaction of the form $|x_i-x_j|^{-k}$ (where $k \to 0$ corresponds to log-gas, while $k=-1$ corresponds to the 1dOCP). Unlike the log-gas, the 1dOCP does not have a determinantal structure and computing its $n$-point correlations would be interesting.

{\it Acknowledgements.} We thank M. Krishnapur, D.~Mukamel, E. Trizac and P. Vivo for discussions. We acknowledge support from the Indo- French Centre for the Promotion of Advanced Research (IFCPAR) under Project 5604-2.

\newpage

\begin{widetext}

\begin{center}\Large{\textbf{Supplementary material}}\end{center}
\vspace*{1cm}

In Eq. (6) of the main text, we have detailed the results for the cumulative distribution $Q(w,N) = {\rm Prob.}(x_{\max}\leq w,N)$. Therefore the corresponding PDF of $x_{\max}$ then reads  
\begin{equation}
P(x_{\max} = w,N) = \partial_w\,Q(w,N) \approx \left\{\begin{array}{rl}
\ee^{-N^3\, \Phi_{-}(w)}  \, &\textrm{for}\,\,0< 2\al - w \sim O(1)\;, \\
    \\
N\,F_\alpha'\left[N(w-2\al) + 2\al \right] \, &\textrm{for}\,\, |2\al-w| \sim 
O(1/N)\\
\\
\ee^{- N^2\, \Phi_+(w)} \, &\textrm{for}\,\,0<w - 2 \al \sim O(1) \;.
\end{array}\right.
\label{sm-result-1}
\end{equation}
In the main text, we have provided a detailed derivation of the central scaling function $F_\alpha(x)$. 
Here we provide the details of the computations of the large deviations functions $\Phi_-(w)$ (left) and $\Phi_+(w)$ (right) respectively. These computations of the large deviations can be carried out along the same line as for the log-gas \cite{sm-Dean06,sm-Dean08,sm-Majumdar09}.

\begin{figure}[hh]
\includegraphics[width = 0.7\linewidth]{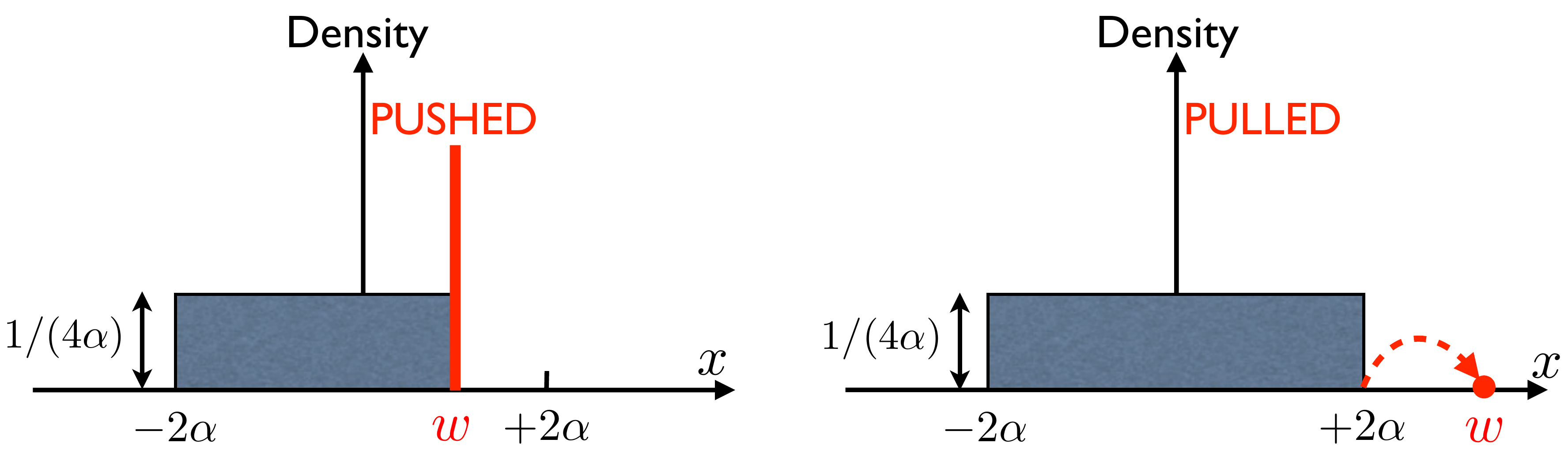}
\caption{{\bf Left:} the left large deviation function $\Phi_-(w)$ [see Eq. (\ref{sm-result-1})] is the free energy cost in {\it pushing} the wall to the left of the right edge, i.e. $w < 2\al$, of the flat equilibrium density. {\bf Right:} the right rate function $\Phi_+(w)$ [see Eq. (\ref{sm-result-1})] is evaluated by computing the energy cost in {\it pulling} a single charge  at $0< w-2\al \sim O(1)$ from the flat equilibrium distribution of charges.}\label{Fig_pushed_pulled}
\end{figure}

\section{Left large deviation function}
To compute the left large deviation function, we start with the cumulative probability $Q(w,N)=\textrm{Prob.}(x_{\max} \le w,N)$ which is given by 
\begin{eqnarray}
Q(w,N)&=& \frac{Z_N(w)}{Z_N(\infty)},~~\text{where,}\label{Q_N-1} \\
Z_N(w)&=& \int_{-\infty}^w dx_1\dots \int_{-\infty}^w dx_N~\ee^{-E(\{x_i\})},~~\text{with} \label{sm-Z_N-w} \\
E(\{x_i\})&=&\frac{N^2}{2} \sum_{i=1}^N x_i^2 - \alpha N \sum_{i \neq j} |x_i-x_j| , \label{sm-E-1dc}
\end{eqnarray}
and $\al > 0$. In the large $N$ limit, the leading behavior of the partition function $Z_N(w)$ of the Coulomb gas with a wall at $w$ can be computed as follows. We first introduce a macroscopic charge density, normalized to unity with its support on $(-\infty,w]$ 
\begin{equation}\label{sm:rho_w}
{\rho}_w(x)= \frac{1}{N} \sum_{i=1}^N \delta(x-x_i) \;.
\end{equation}
The energy of the gas of particles in \eqref{sm-E-1dc}, to leading order for large $N$, can be written in terms of the macroscopic density $\rho_w(x)$ as 
\begin{equation}
E[\{x_i\}]=\mathcal{E}[\rho_w(x)]
\end{equation}
where
\begin{equation}\label{sm-E-Ncube}
\mathcal{E}[\rho_w(x)]=N^3\,\left(\frac{1}{2} \int_{-\infty}^w dx~x^2\rho_w(x) - \al\int_{-\infty}^w dx\int_{-\infty}^w dy~\rho_w(x)\rho_w(y)~|x-y|\right) \;.
\end{equation}
The $N$-fold integration in the partition function $Z_N(w)$ in Eq. (\ref{sm-Z_N-w}) can be carried out in two steps. In the first step, one fixes the macroscopic density $\rho_w(x)$ and one sums over all the microscopic configurations of $x_i$'s consistent with this density $\rho_w(x)$. In the second step, one sums over all possible macroscopic densities $\rho_w(x)$ that are positive and normalized to unity $\int_{-\infty}^w \rho_w(x) \,dx= 1$. The first step gives rise to an entropy term that scales, for large $N$, as $\ee^{O(N)}$. Since this is much smaller than the energy term, which scales like $\ee^{O(N^3)}$ from Eq. (\ref{sm-E-Ncube}), we henceforth neglect the entropy term. Neglecting this entropy term, the partition function can then be expressed as a functional integral    
\begin{equation}
Z_N(w)=\int \mathcal{D}\rho_w ~\text{exp}\left (- \mathcal{E}[\rho_w(x)]\right)\,\delta \left(\int_{-\infty}^w dx~\rho_w(x)-1 \right). \label{partition-2}
\end{equation}
Replacing the delta function by its integral representation, we have 
\begin{eqnarray}
Z_N(w)&=&N^3 \int \frac{d \mu}{2 \pi i} \int \mathcal{D}\rho_w ~\text{exp}\left (- N^3 \,\mathcal{S}[\rho_w(x)] \right), ~~\text{with}\label{partition-3} \\
\mathcal{S}[\rho_w(x)]&=& \left[ \frac{1}{2} \int_{-\infty}^w dx~x^2\rho_w(x) - \al  \int_{-\infty}^w dx\int_{-\infty}^w dy~\rho_w(x)\rho_w(y)~|x-y| +\mu \left(\int_{-\infty}^w dx~\rho_w(x)-1 \right) \right] \;. \label{action}
\end{eqnarray}
For large $N$, we can now evaluate this functional integral by saddle point method. This gives 
\begin{equation}\label{Z_saddle}
Z_N(w) \approx \text{exp}\left (-N^3 S[\rho^*_w(x)] \right),
\end{equation}
where $\rho^*_w(x)$ is the saddle point density that minimizes the action $S[\rho_w(x)]$ in Eq. (\ref{action}). 
The equation for $\rho^*_w(x)$ is obtained from $\left(\frac{\delta S[\rho_w]}{\delta \rho_w(x)}\right)_{\rho_w=\rho^*_w}=0$ as 
\begin{equation}
\frac{1}{2}x^2 -2\alpha \int_{-\infty}^w dy~\rho^*_w(y)~|x-y| +\mu = 0 \;. \label{sadl-eq-1}
\end{equation}
Let us now specify the limits of the integration in the above equation. This equation holds for $x$ belonging to the support of $\rho_w^*(x)$. Clearly the support can not be on $(-\infty,w]$. This can be seen easily from the observation that when $x \to -\infty$, the first term in Eq. (\ref{sadl-eq-1}) scales as $x^2$, while the second term scales as $|x|$ -- hence they can not compensate each other. Therefore the support must be over a finite region $[-B,w]$ where $B$ is to be determined from the normalization condition 
\begin{equation}
\int_{-B}^{w} \rho^*_w(x) dx =1\;. \label{norml-1}
\end{equation}

For $x \in [-B,w]$, it is easy to see by differentiating twice the saddle point equation (\ref{sadl-eq-1}) and using the identity $\frac{d^2}{dx^2}|x-y|=2\delta(x-y)$, that $\rho^*_w(x) = 1/(4 \alpha)$. Clearly, if $w > 2 \al$, the saddle point density is given by 
\begin{eqnarray}\label{rho_star_eq}
\rho_w^*(x) = \frac{1}{4 \al} \; \quad{\rm for} \; \quad-2\al \leq x \leq 2 \al \;, \quad \quad w >  2\al\;.
\end{eqnarray}
Thus for $w > 2\al$ the charge density does not change from its flat equilibrium density -- this is because the charges do not feel the presence of the wall. However, when $w<2\al$, the wall tries to push the charges to the left of $2\al$ (see the left panel of Fig. \ref{Fig_pushed_pulled}). We have seen from above that the bulk density does not change from its equilibrium value $\rho^*_w (x) = 1/(4\al)$ to the left of the wall at $w$. Normalization to unity of the charge density means that the extra charge that the wall displaces must be absorbed at the wall, since the bulk is not affected. This leads, for $w<2\al$, to a new saddle point density of the form   
\begin{equation}
\rho^*_w(x)= \frac{1}{4\al}+ C~\delta(x-w),~\text{for}~-B\le x < w, \label{rho-guess}
\end{equation}
where $C$ represents the density of the charges displaced and absorbed at the wall. The normalization condition $\int_{-B}^w \rho^*_w(x)\,dx = 1$ relates the two parameters $B$ and $C$ via
\begin{equation}
\frac{(w+B)}{4\al} + C=1 \;. \label{eq-3}
\end{equation}
Substituting the saddle point density $\rho^*_w(x)$ in (\ref{sadl-eq-1}) yields
\begin{eqnarray}
\frac{1}{2}x^2 -\frac{1}{2} \int_{-B}^w dy~|x-y|-2 \al~C~ |x-w| +\mu = 0 \;. \label{eq-17-supp}
\end{eqnarray}
Performing the integral over $y$ explicitly, we get
\begin{eqnarray}
\left( 2\al~C + \frac{w-B}{2}\right)x ~+~\left(\mu - 2 \al~C~w-\frac{B^2+w^2}{4}\right)=0, ~~\text{for}~~-B\leq x <w \;. \label{eq-18-supp}
\end{eqnarray}
Since Eq. (\ref{eq-18-supp}) is valid for arbitrary $x \in [-B,w]$, we get two additional equations 
\begin{eqnarray}
&&2\al\,C + \frac{w-B}{2}=0, \label{eq-1}\\ 
&&\mu = 2 \al~C~w+\frac{B^2+w^2}{4}. \label{eq-2}
\end{eqnarray}
We therefore have three equations (\ref{eq-3}), (\ref{eq-1}) and (\ref{eq-2}) for three unknowns $\mu, B$ and $C$. Solving, we get
\begin{eqnarray}
B&=&2\al, \label{B} \\
C&=&\frac{1}{2} - \frac{w}{4\al}, \label{A} \\
\mu &=& \al^2 + \al w - \frac{w^2}{4} \;. \label{mu}
\end{eqnarray}
Note that the condition $C \leq 1$ (needed for the normalization) indicates that the above analysis is only valid for $w > -2\alpha$. When $w \to -2\al$, $C \to 1$: this means that all the charges are absorbed at the wall and there is no bulk charge left. Thus for $w < -2\al$, we have effectively a single charge located at $w$ subjected to a harmonic potential. Therefore, for the saddle point density $\rho_w^*(x)$ we have the following expressions, valid for all $w$  
\begin{eqnarray}\label{rho_summary}
\rho^*_w(x) = 
\begin{cases}
&\frac{1}{4\al}\;,  \quad \hspace*{3.2cm}-2\al \leq x \leq 2\al \;, \quad\quad {\rm for} \quad\qquad\;\;\;\hspace*{0.3cm} w > 2\al \\
& \\
& \frac{1}{4\al} + \left(\frac{1}{2} - \frac{w}{4\al} \right)~\delta(x-w)\;, \quad -2\al\leq x \leq w \;, \quad\quad {\rm for} \quad -2\al \leq w \leq 2\al \\
& \\
&\delta(x-w) \quad \quad \hspace*{5.3cm} {\rm for} \hspace*{1.7cm} w < -2\al \;.
\end{cases}
\end{eqnarray}

Let us first consider $w > 2 \al$. In this case $\rho_w^*(x) = 1/(4 \al)$ for $x \in [-2\al,+2\al]$. Substituting this density in Eq. (\ref{action}) we get the saddle point action
\begin{eqnarray}
S[\rho_w^*(x)] = - \frac{2}{3}\alpha^2\;, \quad \quad {\rm for} \quad w > 2\al \;. \label{S_largew}
\end{eqnarray}
Therefore from Eq. (\ref{Z_saddle}) the partition function $Z_N(w)$ for large $N$ and for $w > 2\al$ behaves as 
\begin{eqnarray}
Z_N(w) \approx \ee^{\frac{2}{3}\alpha^2\,N^3 } \;, \quad \quad {\rm for} \quad w > 2\al \;. \label{Z_largew}
\end{eqnarray}
In particular, taking $w \to \infty$ limit, we obtain the denominator in Eq. (\ref{Q_N-1}) as
\begin{eqnarray}\label{denominator}
Z_N(\infty) \approx \ee^{\frac{2}{3}\alpha^2\,N^3} \;.
\end{eqnarray}
Hence, finally, for $w > 2\al$, to leading order for large $N$, we get
\begin{eqnarray}\label{Q_largew}
Q(w,N) = \frac{Z_N(w)}{Z_N(\infty)} \approx 1  \;, \quad \quad {\rm for} \quad w > 2\al \;.
\end{eqnarray}
To calculate the corrections to this leading order result, we need to consider the right large deviations function, that will be 
computed in the next section.  

Let us know consider the region where $-2\al \leq w \leq 2\al$. Substituting the saddle point density $\rho_w^*(x)$ from the second line of Eq.~(\ref{rho_summary}) in Eq. (\ref{action}) we get
\begin{eqnarray}\label{S_inter}
S[\rho_w^*(x)] = - \frac{8\al^3+12\al^2w-6\al w^2+w^3}{24 \al} \;, \quad \quad {\rm for} \quad -2\al \leq w \leq 2\al \;.
\end{eqnarray}
Substituting this result in Eq. (\ref{Z_saddle}) and using the expression for the denominator in Eq. (\ref{denominator}) 
we get
\begin{eqnarray}\label{Q_mediumw}
Q(w,N) = \frac{Z_N(w)}{Z_N(\infty)} \approx \ee^{-N^3\Phi_-(w)}  \;, \quad {\rm where} \quad \Phi_-(w) = \frac{(2\al - w)^3}{24\,\al}\;, \quad \quad {\rm valid \quad for} \quad  -2\al \leq w \leq 2\al \;.
\end{eqnarray}

Finally, we consider the region where $w\leq-2\al$. Substituting the saddle point density $\rho_w^*(x)$ from the third line of Eq.~(\ref{rho_summary}) in Eq. (\ref{action}) we get
\begin{eqnarray}\label{S_small}
S[\rho_w^*(x)] = \frac{w^2}{2}\;, \quad \quad {\rm for} \quad w < -2\al \;.
\end{eqnarray}
Substituting this result in Eq. (\ref{Z_saddle}) and using the expression for the denominator in Eq. (\ref{denominator}) 
we get
\begin{eqnarray}\label{Q_smallw}
Q(w,N) = \frac{Z_N(w)}{Z_N(\infty)} \approx \ee^{-N^3\Phi_-(w)}  \;, \quad {\rm where} \quad \Phi_-(w) = \frac{w^2}{2} + \frac{2}{3}\alpha^2\;, \quad \quad {\rm valid \quad for} \quad  w\leq -2\al\;.
\end{eqnarray}
Summarizing we obtain the result in Eq. (9) of the main text
\begin{eqnarray}\label{sm-left_rf}
\Phi_{-}(w) = 
\begin{cases}
& \frac{(2\al-w)^3}{24\al}\;,\;\quad -2\al\leq w \leq 2\al \\
&Ê\vspace*{-0.3cm} \\
& \frac{w^2}{2} + \frac{2}{3} \al^2 \;, \;\quad\; w \leq -2\al \;.
\end{cases}
\end{eqnarray}
In Fig. \ref{plot_phiminus}, we show a plot of $\Phi_-(w)$ as a function of $w$.

{\it Third order phase transition at $w = 2\al$.} The cumulative distribution $Q(w,N)$ in Eq. (\ref{Q_N-1}) is the ratio of two partition functions. Hence $-\ln Q(w,N) = -\ln Z_N(w)+\ln Z_N(\infty)$ can be interpreted as a free energy difference. Indeed, from Eq.~(\ref{Q_largew}), we see that to leading order for large $N$, $-\ln Q(w,N) \approx 0$ for $w>2\al$. In contrast, for $w<2\al$, using Eq. ~(\ref{Q_mediumw}) and (\ref{Q_smallw}), we see that $-\ln Q(w,N) \approx N^3 \Phi_-(w)$ where $\Phi_-(w)$ is given in Eq. (\ref{sm-left_rf}). Hence, we get (see Fig. \ref{plot_phiminus})
\begin{eqnarray}\label{log_Qn}
- \lim_{N \to \infty} \frac{\ln Q(w,N)}{N^3} =
\begin{cases}
&0 \;, \;\hspace*{0.85cm} {\rm for}\; w > 2\al \\
&\Phi_-(w) \;, \; {\rm for}\; w < 2\al \;.
\end{cases}
\end{eqnarray}

Thus $\Phi_-(w)$ is just the free energy cost in pushing the wall $w$ to the left of the right edge $2 \al$ (see Fig. \ref{Fig_pushed_pulled}). From the expression of $\Phi_-(w)$ in the first line of Eq. (\ref{sm-left_rf}), it follows that $\Phi_-(w)$ vanishes as the third power $\Phi_-(w) \propto (2\al-w)^3$ as $w \to 2\al$ from the left. Thus the third derivative of the free energy difference vanishes at the critical point $2 \al$, making this a third order phase transition. Indeed, one can also look at the pressure on the wall, which is simply ${\cal P} = -N^3\,\Phi'_-(w)$ (derivative of the free energy with respect to the wall position). Clearly, the pressure ${\cal P}$ is zero for $w > 2\al$ (the charges do not touch the wall) and is non zero for $w<2\al$. The pressure ${\cal P}$ vanishes as $N^3\,(2\al - w)^2$ as $w \to 2\al$ from below.

\begin{figure}
\includegraphics[width = 0.7\linewidth]{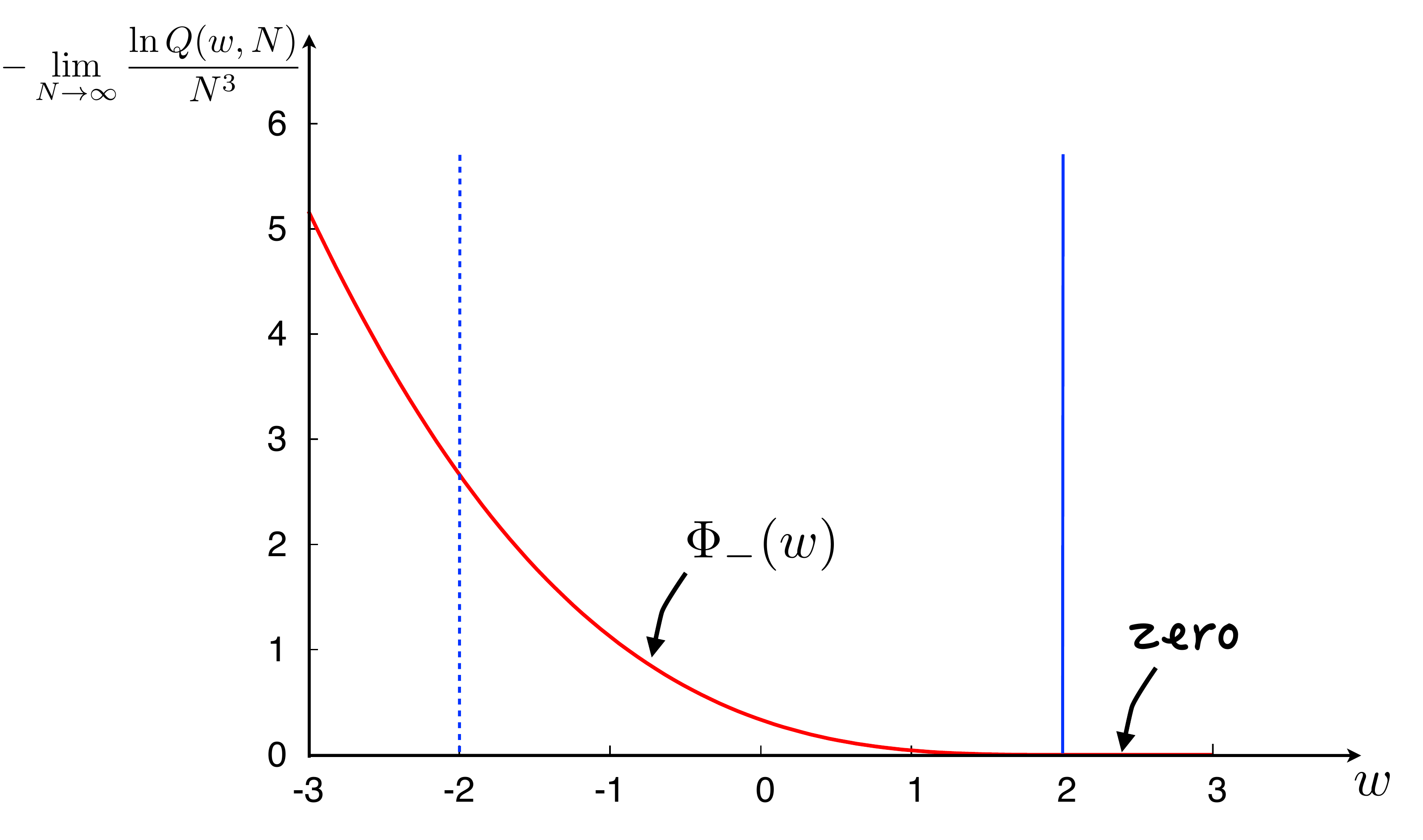}
\caption{Plot of the free energy $-\lim_{N \to \infty} {\ln Q(w,N)}/{N^3}$ as a function of $w$ for $\al=1$. For $w>2\al = 2$, the limiting value is just zero, $-\lim_{N \to \infty} {\ln Q(w,N)}/{N^3} = 0$, while it is non-zero for $w< 2 \al$, $-\lim_{N \to \infty} {\ln Q(w,N)}/{N^3} = \Phi_-(w)$ [see Eq. (\ref{log_Qn})]. This transition at $w = 2\al$ is depicted by the solid blue line 
in Eq. (\ref{log_Qn}). Given that $\Phi_-(w) \propto (2- w)^3$ as $w \to 2$ [see Eq. (\ref{sm-left_rf})], this is a third order phase transition. The dotted blue line at $w = - 2\al = -2$ indicates a second third order transition that occurs in this model [see Eq. (\ref{sm-left_rf})] when the pushed gas behaves as a single charged particle located at $w$ [see Eq. (\ref{rho_summary})].}\label{plot_phiminus}
\end{figure}

\section{Right large deviation function}

Here we focus, for large $N$, on the distribution $Q(w,N)$ in the region $0<w-2\al \sim O(1)$. From the analysis in the previous section,  we have seen that in this regime, to leading order for large $N$, $Q(w,N) \approx 1$ [see Eq. (\ref{Q_largew})]. To compute the subleading corrections to this leading order term $1$, it is convenient to consider instead the PDF of $x_{\max}$, given by the derivative of Eq. (\ref{Q_N-1}) 
\begin{equation}\label{exact_PDF}
P(w,N) = \partial_w Q(w,N) = \frac{N}{Z_N(\infty)} \ee^{-\frac{N^2}{2}w^2} \int_{-\infty}^w dx_1 \cdots \int_{-\infty}^w dx_{N-1}\;\ee^{- \alpha\,N \sum_{j=1}^N |w-x_j| - \al\,N \sum_{1\leq i\neq j\leq N-1}|x_i-x_j| - \frac{N^2}{2}\sum_{i=1}^{N-1}x_i^2} 
\end{equation}
where we have set $x_N = w$ in Eqs. (\ref{sm-Z_N-w}) and (\ref{sm-E-1dc}). This can be re-written as
\begin{eqnarray}\label{PDF_largeN_1}
P(w,N) = \frac{N\,Z_{N-1}(\infty)}{Z_N(\infty)}\, \ee^{-\frac{N^2}{2}w^2} \Big \langle \ee^{-2\al \,N \sum_{j=1}^{N-1}(w-x_j)}\Big\rangle_{N-1} \;,
\end{eqnarray}
where $\langle \cdots \rangle_{N-1}$ denotes the average over the $N-1$ charges. We can then analyse this average for large $N$, for $w > 2 \al$, following Ref. \cite{sm-Majumdar09} for the log-gas in the corresponding right large deviation regime. To evaluate this average, we note that essentially one single charge out of $N$ is detached at $w > 2\al$, while the rest of $N-1$ charges should be in their equilibrium flat configuration, i.e., with a density $\rho^*_w(x) = 1/(4 \al)$ for $x \in [-2\al,2\al]$ (see the right panel of Fig. \ref{Fig_pushed_pulled}). Furthermore, for large $N$, to leading order, we can approximate  
\begin{eqnarray}\label{PDF_largeN_2}
P(w,N) \approx \ee^{-\frac{N^2}{2}w^2 - 2 \al \, N \langle \sum_{j=1}^{N-1} (w-x_j)\rangle} \approx \ee^{-N^2\left(\frac{w^2}{2}-2\al \int_{-2\al}^{2\al} (w-x)\rho^*_w(x) dx + C_0\right)} \;,
\end{eqnarray}
where $C_0$ is a constant, i.e. independent of $w$. Using $\rho^*_w(x) = 1/(4 \al)$ for $x \in [-2\al,2\al]$ and performing the integral in Eq. (\ref{PDF_largeN_2}), we obtain 
\begin{eqnarray}\label{sm_Epulled}
P(w,N) \approx \ee^{-\Delta E_{\rm pulled}} \approx \ee^{-N^2\,  \Phi_+(w)} \;,
\end{eqnarray}
where
\begin{eqnarray}\label{sm-phi+}
\Phi_+(w) = \frac{(w-2\al)^2}{2} \;, \quad w > 2\al \;.
\end{eqnarray}
Thus $\Delta E_{\rm pulled}$ in Eq. (\ref{sm_Epulled}) corresponds to the energy in pulling out a single charge from the equilibrium configuration of charges with a flat density.

\end{widetext}

\end{document}